\documentclass[epj,nopacs]{svjour}
\usepackage{epsfig}

\newcommand{\munp}{\mbox{$\mu$-$\nu$~}}

\newcommand{\mujnp}{\mbox{$\mu$-$j$-$\nu$~}}

\newcommand{\jjjj}{\mbox{$j$-$j$-$j$-$j$~}}

\newcommand{\ejjjj}{\mbox{$e$-$j$-$j$-$j$-$j$~}}

\begin{document}
\title{General Search for New Phenomena in \boldmath{$ep$} Scattering at HERA}
\author{Martin Wessels\inst{1}\thanks{on behalf of the H1 Collaboration}}
\institute{\inst{1} I. Physikalisches Institut der RWTH, Aachen, Germany}
%
\date{Received: date / Revised version: date}
\abstract{
  A model-independent search for deviations from the Standard Model 
  prediction has been performed in $e^+ p$ and $e^- p$ collisions 
  at HERA using H1 data. All experimentally measurable event topologies
  involving isolated electrons, photons, muons, neutrinos and jets with
  high transverse momenta have been investigated.
  A good agreement with the Standard Model prediction is found 
  in most of the event classes. A new algorithm has been developed to look 
  for regions with large deviations from the Standard Model in the invariant 
  mass and sum of transverse momenta distributions and to quantify the
  significance of the fluctuations observed. 
  The largest deviation is found in topologies with an isolated muon,
  missing transverse momentum and a jet which confirms previous observations.
  About $2\%$ of hypothetical Monte Carlo 
  experiments would produce deviations more significant than the
  one observed in the corresponding distribution of sum of transverse momenta.
  }
\titlerunning{General Search for New Phenomena in $ep$ Scattering at HERA}
\maketitle
\section{Introduction}
\label{sec:intro}
At HERA electrons\footnote{
  In this paper "electron" refers to both electrons and positrons, if
  not otherwise stated.}
and protons collide with a centre-of-mass energy of up to $319$~GeV.
The H1 experiment at HERA has accumulated data corresponding to more
than $100$~$\mbox{pb}^{-1}$ of integrated luminosity in
the first period (HERA I, 1994-2000) and provides 
therewith a complete and well understood data set.
One important goal at HERA is the search for new physics beyond the Standard 
Model (SM), which is predicted by a large variety of extensions to the SM
resulting in final state topologies at high energies or large transverse momenta.
In various dedicated analyses the HERA I data have been searched for new 
physics signals, and upper limits on cross-sections of new processes have been derived. 
Some discrepancies between the analysed data and the SM prediction have been 
found \cite{ref:isolep,ref:dielec}.\\[.2cm]
The approach described in this analysis consists in an extensive search for 
deviations from the SM prediction at large transverse momentum $P_T$ in all 
final state topologies with at least two objects. The analysis covers 
phase space regions where the SM prediction is sufficiently precise to detect 
anomalies and does not rely on assumptions concerning the characteristics of specific 
models beyond the SM. An in this spirit so-called model-independent search might 
therefore be able to discover unexpected manifestations of new physics and give 
an answer to the important question if new physics signals might be hidden 
in the HERA I data.
\begin{figure}[t]
  \begin{center}
    \resizebox{0.46\textwidth}{!}{\epsfig{file=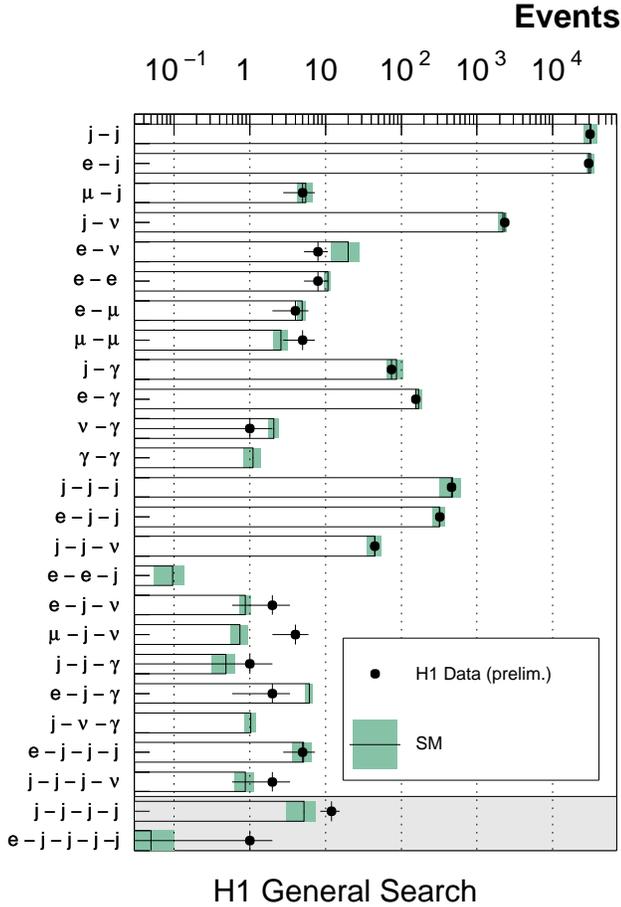,angle=270,clip=}}
    \caption{
      The data and SM expectation for all event classes containing data events or 
      a SM expectation greater than $0.1$ event. The predictions for the
      \jjjj and \ejjjj event class (grey area) are less reliable, and 
      these classes are therefore not passed through the statistical analysis.
      }
  \end{center}
  \label{fig:sumplot}
\end{figure}
\section{Data analysis}
\label{sec:data}
The event sample studied consists of the full 1994-2000 HERA I data.
Investigated are all final states with at least two objects with $P_T>20$~GeV 
in the polar angle range \mbox{$10^\circ < \theta < 140^\circ$}. 
The considered objects are electron ($e$), muon ($\mu$), photon ($\gamma$), jet ($j$) 
and neutrino ($\nu$) (or non-interacting particle). The identification criteria for 
each type of particle are based on previous analyses on specific final states. 
Additional requirements have been chosen to ensure an unambiguous identification 
of particles, still keeping high efficiencies \cite{ref:eps118}. 
Moreover all objects are required to be isolated from each other by a minimum distance 
\mbox{$R=\sqrt{\Delta\eta^2+\Delta\phi^2}>1$} in the \mbox{$\eta-\phi$} plane.
All events are then divided in exclusive event classes according to the
number and type of the measured objects. This exclusive classification ensures a clear 
separation of the final states and an unambiguous statistical interpretation later on.\\
As this analysis investigates all final state topologies of $ep$ interactions, 
a precise and reliable estimate of all relevant HERA processes is needed. Hence, several 
Monte Carlo generators are used to generate a large number of events in all event classes,
carefully avoiding double-counting of processes. The simulation contains the order 
$\alpha_S$ matrix elements for QCD processes, while second order $\alpha$ matrix elements
are used to calculate QED processes. Additional jets are modeled using leading logarithmic 
parton showers as representation of higher order QCD radiation.
All processes are generated with a 
luminosity at least 20 times higher than that of the data.\\[0.2cm]
The results of the analysis are summarised in Figure~\ref{fig:sumplot}, which presents 
the event yields for the data and SM expectation subdivided in event classes. Shown are all 
event classes containing data events or a SM expectation greater than 0.1 event\footnote{
  The \munp event class was discarded from the analysis because of overwhelming background 
  from low $P_T$ photoproduction.}.\\
A good overall agreement between data and SM expectation is observed for most of the event 
classes. Many of them have been analysed herein for the first time at HERA.
Selection efficiencies have been derived to quantify the finding potential and can be 
used to set exclusion limits for new physics signals~\cite{ref:eps118}.\\ 
A discrepancy between data and SM expectation is observed in the \mujnp event class, which 
corresponds to typical event topologies arising from $W$ production. The deviation was already 
reported in~\cite{ref:isolep} and will be further discussed in section~\ref{sec:search}.\\
Some discrepancies on the total event yields can also be observed in the \jjjj and \ejjjj
event classes, but since these spectacular events can -- in the current Monte Carlo programs -- 
only be produced via parton showers, it can not be ensured that the Monte Carlo prediction is reliable.
\begin{figure}[b]
  \begin{center}
    \resizebox{0.41\textwidth}{!}{\epsfig{file=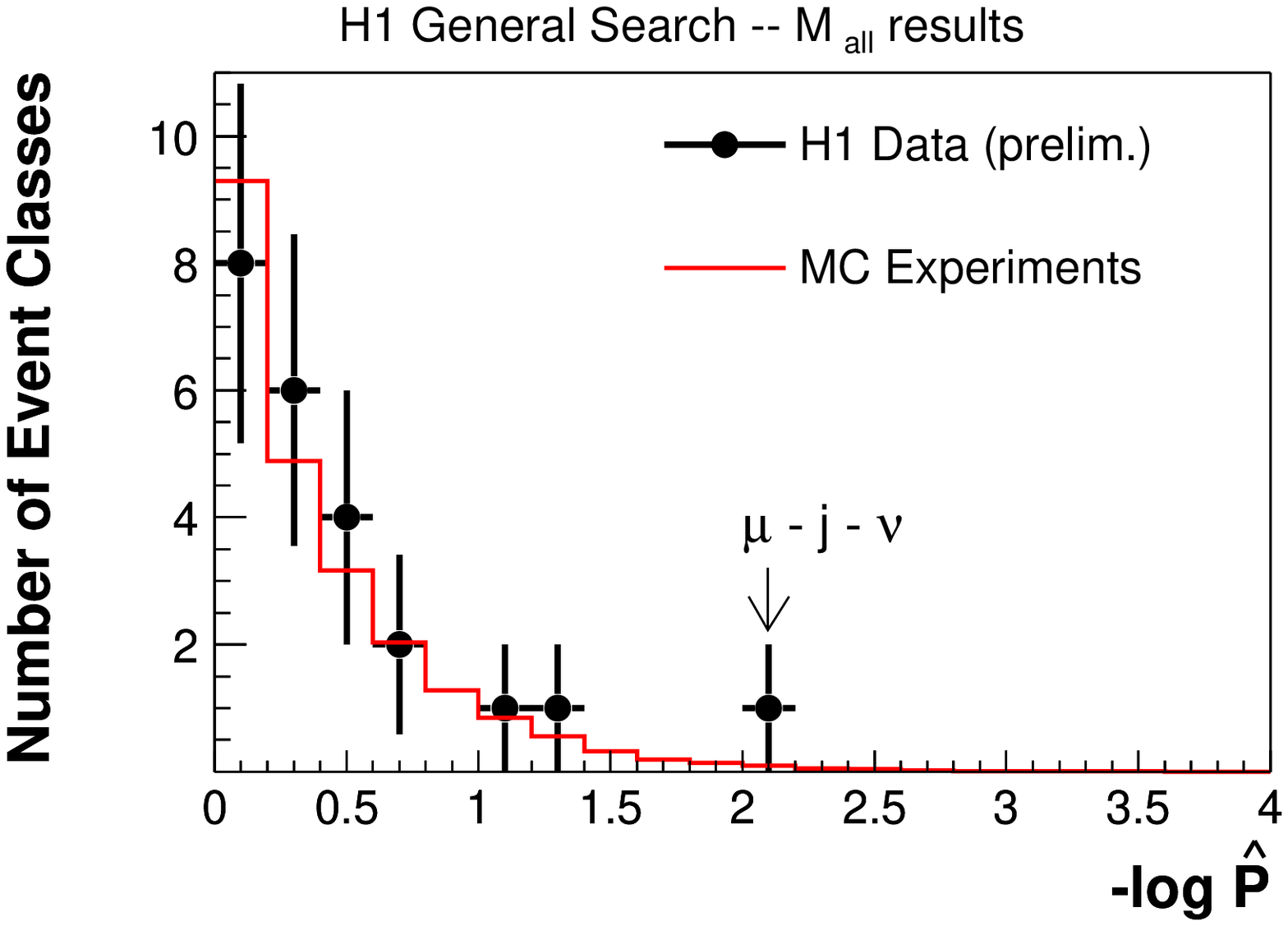,clip=}}
    \resizebox{0.41\textwidth}{!}{\epsfig{file=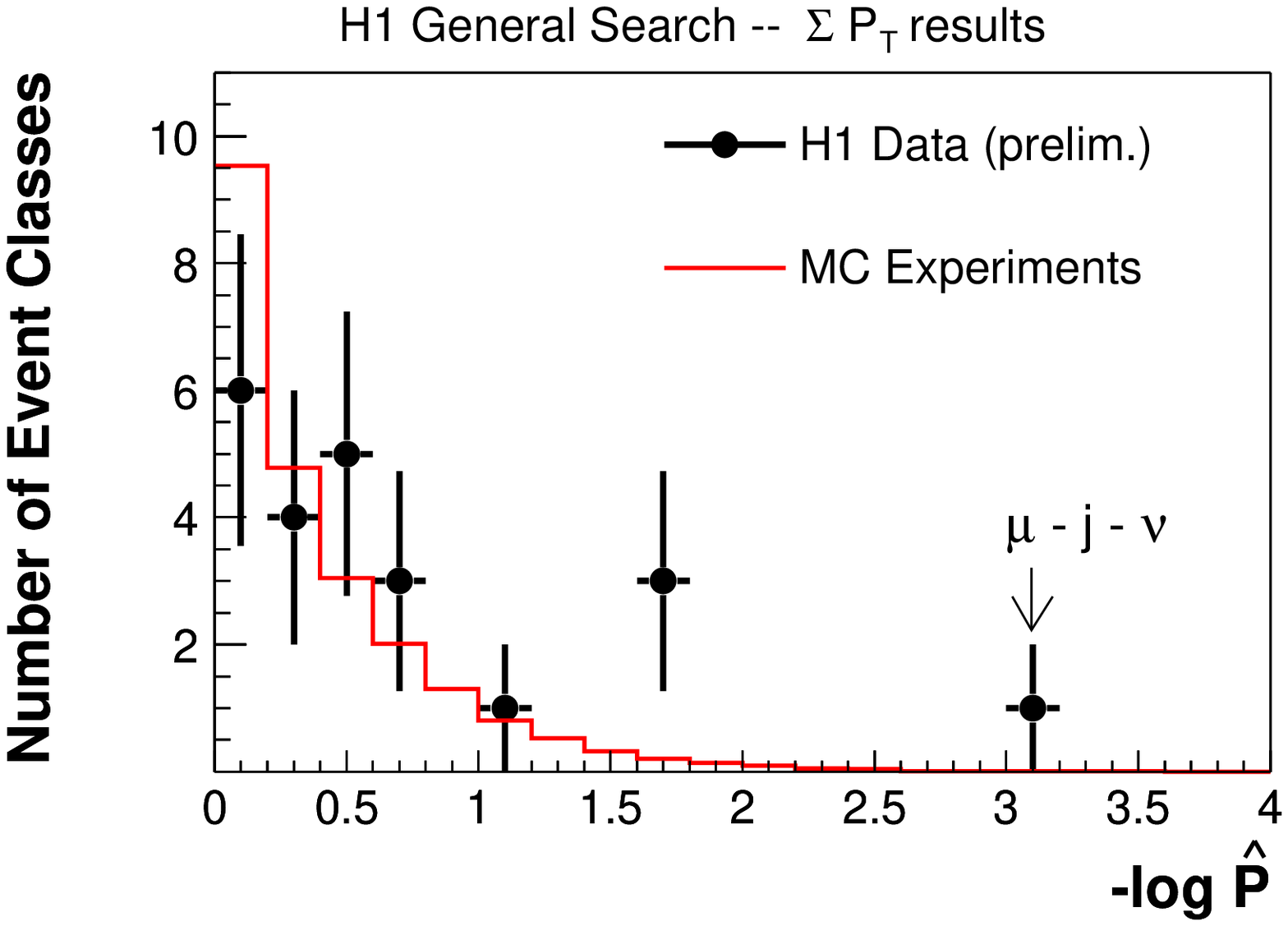,clip=}}
    \caption{
      The $-\log{\hat{P}}$ values for the data event classes and the
      expected distribution from MC experiments for the search in the 
      $M_{\rm all}$ (upper) and in the $\sum P_T$ 
      distributions (lower). All event classes with a SM expectation 
      greater than $0.1$ event, except the \jjjj and the \ejjjj event class, 
      are presented.}
  \end{center}
  \label{fig:phat}
\end{figure}
\section{Search for deviations}
\label{sec:search}
\begin{figure}[t]
  \begin{center}
    \resizebox{0.39\textwidth}{!}{\epsfig{file=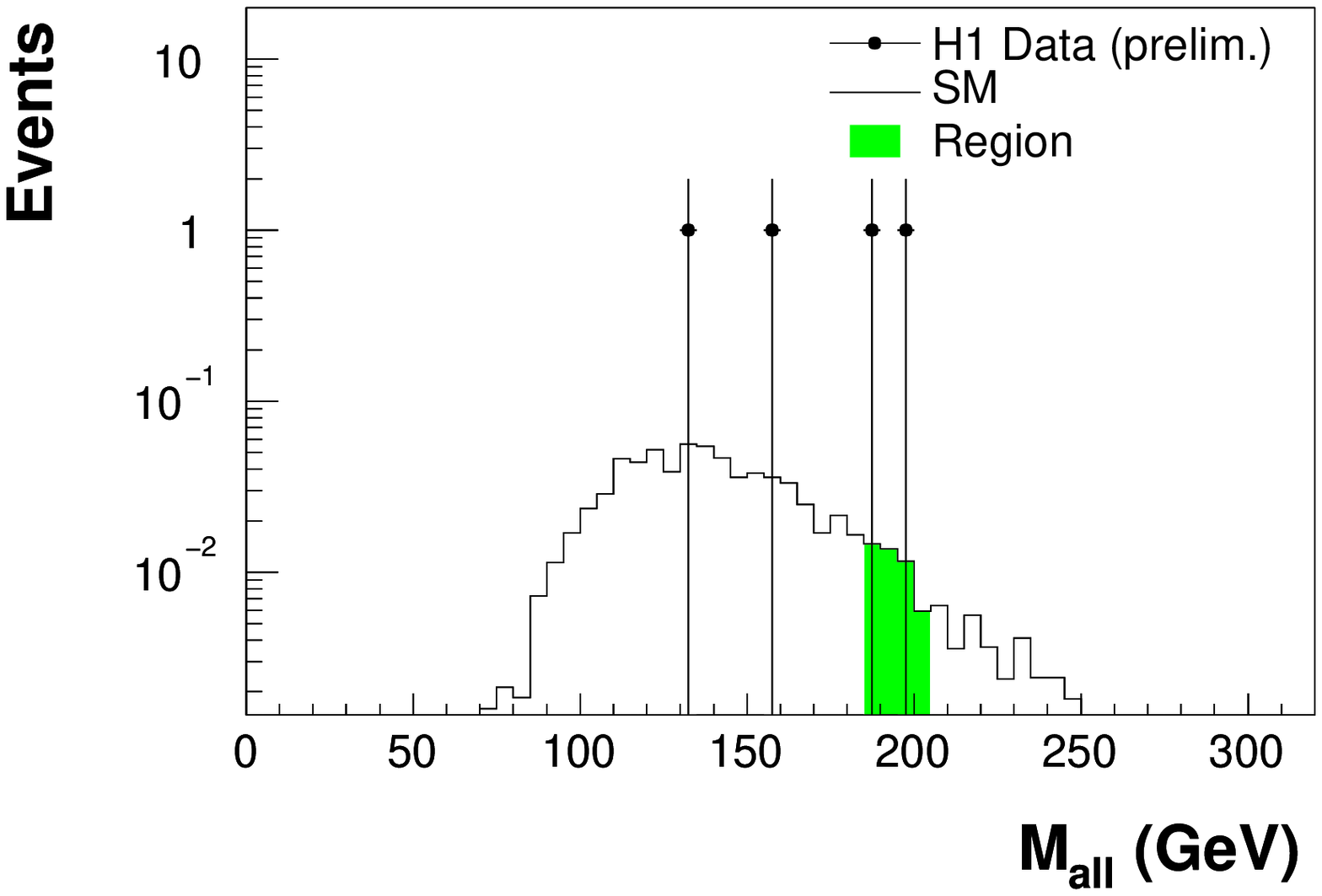,clip=}}
    \resizebox{0.39\textwidth}{!}{\epsfig{file=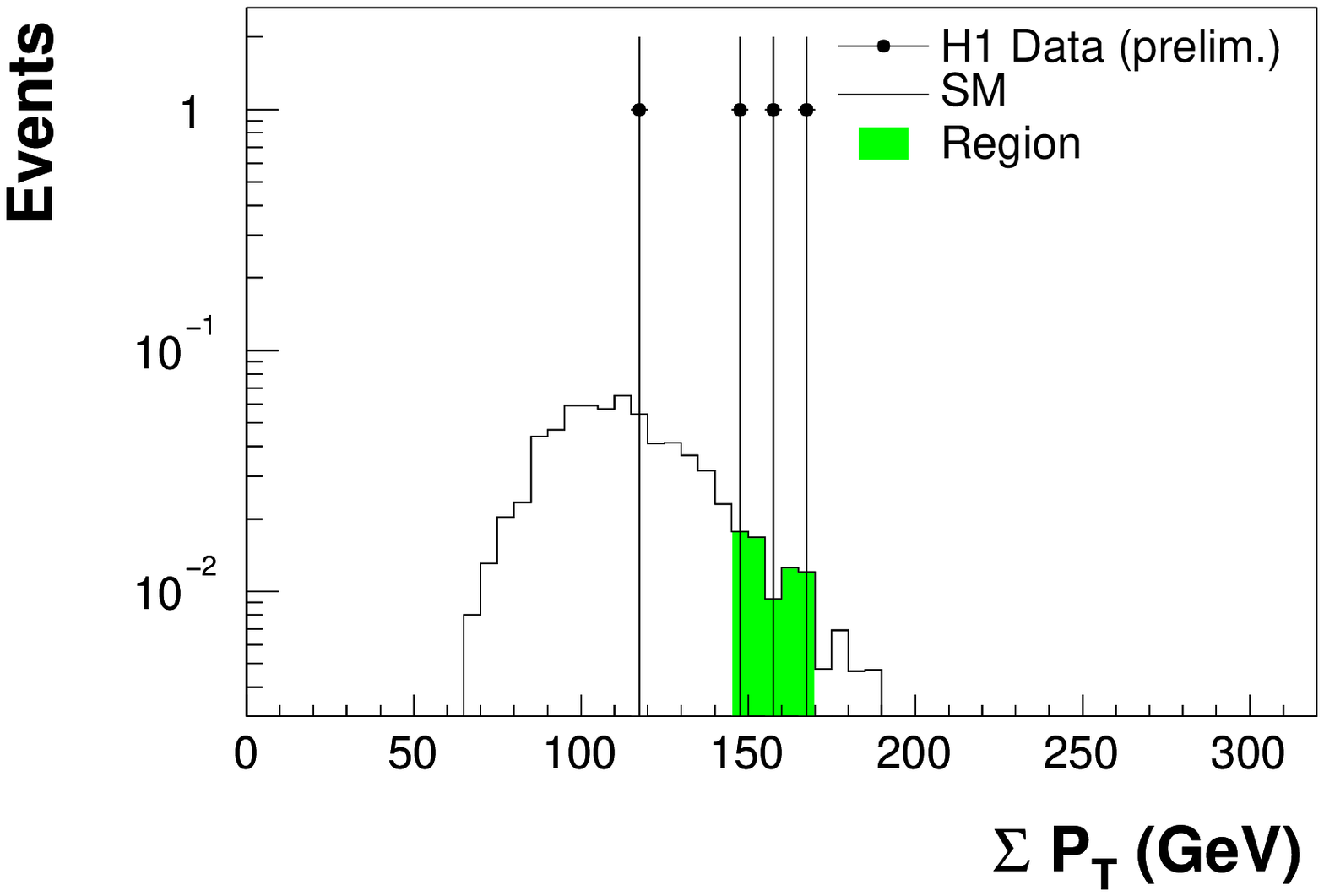,clip=}}
    \caption{
      The number of data events and the SM expectation for the \mujnp event class 
      as a function of $M_{\rm all}$ (upper) and $\sum{P_T}$ (lower). 
      The shaded area shows the region of greatest deviation 
      chosen by the search algorithm.}
  \end{center}
  \label{fig:regions}
\end{figure}

To search for deviations between data and the SM expectation the invariant mass $M_{\rm all}$ 
and sum of transverse momenta $\sum P_T$ distributions of all reliable event classes 
are investigated. In order to quantitatively determine the level of agreement between data 
and SM expectation and to identify regions of possible deviations, a new
search algorithm has been developed. A region is defined as a sample of connected 
histogram bins, which have at least the size of twice the resolution of the observable. 
A statistical estimator $p$ is defined to determine the region of most interest
by calculating the probability, that the SM expectation fluctuates upwards or 
downwards to the data. This estimator is derived from the convolution of a Poisson probability 
density function (pdf) to account for statistical errors with a Gaussian pdf 
to include systematic uncertainties~\cite{ref:eps118}.
As in this ansatz a possible sign of new physics is found, if the expectation significantly 
disagrees with the data, the region of most interest (greatest deviation) is given by the 
region having the smallest $p$-value, $p_{\rm min}$. This method finds narrow 
resonances, single outstanding events as well as signals spread over large regions of 
phase space in distributions of any shape.\\
The fact that somewhere in the distribution a fluctuation with a value $p_{\rm min}$
might occur is taken into account by calculating the probability $\hat{P}$, to observe a deviation 
with a $p$-value $p_{\rm min}$ at any position in the distribution.
Thus $\hat{P}$ is the central measure of significance of the found deviation. 
The calculation of the global significance of a deviation has been inspired by \cite{ref:Abbott:2000fb}. 
To determine $\hat{P}$ hypothetical data histograms are diced according to the probability 
density function of the expectation. The value of $\hat{P}$ is then defined as the fraction of 
hypothetical data histograms with a $p_{\rm min}$-value smaller than the $p_{\rm min}$-value 
obtained with the real data,
and consequently the event class of most interest for a search is the one
with the smallest $\hat{P}$-value.\\
To compare the obtained $\hat{P}$-values with an expectation, all data distributions are
replaced by hypothetical Monte Carlo (MC) distributions. The complete algorithm is applied 
on these independent sets of MC experiments. In the case that deviations arise only from
statistical or systematical fluctuations, the distribution of $\hat{P}$-values obtained from
data and MC experiments are compatible.\\[0.2cm]
The results of the search for deviations between data and SM expectation are summarised in
Figure \ref{fig:phat}. Presented are the distributions of 
the negative decade logarithm of the final $\hat{P}$-values obtained from data compared to the 
expectation from MC experiments. The upper figure shows the distribution obtained 
from the search in the $M_{\rm all}$ distributions, while the result of the search in the 
$\sum{P_T}$ distributions is presented in the lower figure. Most $\hat{P}$-values
range from 0.01 to 0.99, corresponding to event classes where no significant discrepancy between 
data and SM expectation  is observed.\\
The largest deviation is observed in the \mujnp event class, where 
$\hat{P}$-values of 0.010 and 0.0008 are found corresponding to the high $M_{\rm all}$
and high $\sum{P_T}$ region. The corresponding distributions of $M_{\rm all}$ and 
$\sum{P_T}$ together with the regions selected by the algorithm are presented in 
Figure \ref{fig:regions}. The invariant mass region contains 2 data events for an expectation of 
0.05$\pm$0.02 events. In the chosen $\sum{P_T}$ region 3 data events are found while only 
0.07$\pm$0.03 events are expected. This discrepancy was already reported in~\cite{ref:isolep}.\\
As this analysis studies a large number of event classes, there is some chance that 
small $\hat{P}$-values can arise. To quantify the significance of the deviations, the 
likeliness can be calculated, that the smallest $\hat{P}$-value found in the investigated
$M_{\rm all}$ and $\sum P_T$ distributions occurs. These values are found to be about
25\% for the set of $M_{\rm all}$ distributions and about 2\% for the $\sum P_T$ 
distributions.
\section{Conclusions}
\label{sec:concl}
The data collected with the H1 experiment during the years 
1994-2000 (HERA I) has been searched for deviations from the SM prediction.
All possible event topologies have been investigated 
in a coherent and model-indepen\-dent way. Many event classes are analysed herein 
for the first time at HERA. A good agreement between data and SM expectation 
has been found in most event classes, illustrating the good understanding of 
SM physics at HERA up to the edges of phase space. 
The invariant mass and sum of transverse momenta 
distributions of the event clas\-ses have been systematically searched for 
deviations with a novel algorithm. 
The most significant deviation is found in the \mujnp event class, 
a topology, where deviations have also been previously observed. 
No new significant deviation is found.
With this work one of the most complete analysis of HEP data at high $P_T$
has been presented and we are curiously looking forward to HERA II data taking.
\section*{Acknowledgements}
\label{sec:acknow}
I gratefully acknowledge my collaborators in this work, 
Sascha Caron, Gilles Frising, Matti Peez and Emmanuel Sauvan.

\begin{thebibliography}{9}
\bibitem{ref:isolep}
  V.~Andreev {\it et al.} [H1 Collaboration],  
  Phys.\ Lett.\ B {\bf 561} (2003) 241, arXiv:hep-ex/0301030.
\bibitem{ref:dielec}
  A.~Aktas {\it et al.} [H1 Collaboration],
  Eur.\ Phys.\ J.\ C {\bf 31} (2003) 17, arXiv:hep-ex/0307015.  
\bibitem{ref:eps118}
  H1 Collaboration, paper 118, submitted to EPS-HEP 2003.
\bibitem{ref:Abbott:2000fb}
  B.~Abbott {\it et al.} [D0 Collaboration], 
  Phys.\ Rev.\ D {\bf 62} (2000) 092004, arXiv:hep-ex/0006011.
\end{thebibliography}
\end{document}